The Fifth Information Systems International Conference 2019

# Why does cultural diversity foster technology-enabled intergenerational collaboration?

Irawan Nurhas[a,c,]*, Bayu Rima Aditya[b], Stefan Geisler[a], Jan Pawlowski[a,c]

[a]*Institute of Positive Computing, Hochschule Ruhr West University of Applied Sciences, Bottrop, Germany*
[b]*School of Applied Science, Telkom University, Bandung, Indonesia*
[c]*Faculty of Information Technology, University of Jyväskylä, Jyväskylä, Finland*

**Abstract**

Globalization and information technology enable people to join the movement of global citizenship and work without borders. However, different type of barriers existed that could affect collaboration in today's work environment, in which different generations are involved. Although researchers have identified several technical barriers to intergenerational collaboration (iGOAL), the influence of cultural diversity on iGOAL has rarely been studied. Therefore, using a quantitative study approach, this paper investigates the impact of differences in cultural background on perceived technical and operational barriers to iGOAL. Our study reveals six barriers to IGC that are perceived differently by culturally diverse people (CDP) and non-CDP. Furthermore, CDP can foster IGC because CDP consider the barriers to be of less of a reason to avoid working with different generations than do non-CDP.

*Keywords:* challenges; problems; cross-generational collaboration; cross-cultural teamwork; barriers

---

* Corresponding author. Tel.: +49-208-88254-853.
  *E-mail address:* irawan.nurhas@hs-ruhrwest.de

## 1. Introduction

The global citizen is an upcoming force in the globalization phenomenon, which encourages mobility between countries or continents and simultaneously promotes collaborative innovation based on diversity (e.g., gender, age, nationality, or culture). More importantly, with recent advances in information technology, globalization offers the possibility for collaboration among individuals of different ages [8, 17]. Understanding how to integrate older people into the innovation process is therefore an emerging issue [15, 42]. Intergenerational collaboration (iGOAL) can accelerate innovation, which is considered to be one of the most important success factors for sustainable family businesses [13, 25, 29] and global enterprises [22, 42]. Moreover, iGOAL also supports innovation through the exchange of knowledge and skills [18, 25]. Age and cultural differences in the workplace can create various barriers and challenges for group work, communication practices, and technology use [15, 24, 25]. In this study, we define barriers as all conditions that can prevent an individual from working with another individual when the age gap between the two is large. In addition to the existing barriers to and challenges presented by collaborative innovation, in the coming decades, companies will face demographic changes [39] to integrate older generations and strengthen iGOAL [15, 17, 42].

The challenge of developing a collaborative system to overcome barriers due to age differences in global collaboration remains largely unexplored, especially for CDP [16, 17, 27]. Indeed, CDP are global citizens who will use the systems in place in teams that comprise members from different generations [17, 22]. In this study, we focus on examining barriers in a global context (cultural diversity based on different nations or continents). Therefore, we coined the term CDP for people with a different cultural background than the native population in the areas in which they live [27]. In the context of global innovation, for example, we work with experienced older adults to support young entrepreneurs in an international context. In other words, older adults with strong expertise in international markets can provide younger generations with vast knowledge and support to run a business in a global environment while receiving gratitude in return and encouraging the participation of society. Researchers from different fields have identified several technical barriers to iGOAL [24, 25] and global collaboration [32, 35]. However, these studies did not integrate CDP, a practice of lifestyle for example as a digital nomad, which is gaining momentum worldwide [12, 15, 16]. Indeed, CDP are fueling global innovation [15], and ignorance of the correlated barriers will have negative impact on organizations in the innovation process [15, 17]. Therefore, we conducted a quantitative study [21, 43] in a global context to compare CDP and non-CDP audiences. We developed a hypothesis based on the literature on technical barriers in iGOAL. We then evaluated the perception of CDP and non-CDP by applying a Chi-square test and a t-test [26, 37] to our hypothesis to uncover which barriers to iGOAL correlate with cultural differences. The results of this study provide an understanding of the system requirements to support cultural differences in the iGOAL environment, as well as an overview of why can the CDP support the intergenerational innovation process.

This paper consists of the following: First, we present our hypothesis based on the literature on technical barriers related to iGOAL. Subsequently, the methodology used to evaluate the barriers and the hypothesis is presented in the third section. Afterward, we present the results of the study, followed by analysis and discussion. Finally, we provide the conclusions of the study and the future research direction.

## 2. System Design Approach and the Hypothesis Development

Significant age-gap disparities can create multiple dimensions of barriers [3, 5, 25] and lead to differences in the adoption of technologies to support collaboration [15, 32]. However, in this study, we focused on the technical and operational dimensions [5, 32] of barriers experienced by individuals with different cultural backgrounds in iGOAL. We specifically considered the technical and operational dimensions because the analysis results of the technical barriers serve as a starting point for the system designer to identify the requirements of the system, and thus enable technology-based multicultural collaboration [32] in an intergenerational context. This section therefore proposes a hypothesis and discusses a number of barriers related to the technical and operational dimensions of iGOAL. In regard to age differences, the age factor is not only a numerical issue, it is also, and more importantly, an issue related to differences in experience (DTF) [14, 15, 19]. Furthermore, an unsupportive technological environment (UTE) can emerge [23, 41], as each country has different technological infrastructures [6, 32] that are perceived differently over time and based on culture [11]. Indeed, a known technology for a particular generation or country could be a new technology for others [6]. Adaptation to new environmental technologies may be more difficult since the environment differs greatly depending on the current skills and experience of the generation [3, 6]. Moreover, a lack of training in digital technology for collaboration purposes can also be a problem, not only for another generation but also for another culture (TDC) [6, 11].

Collaboration involves a commitment on both sides, including the commitment to invest one's time [3]. However, due to the different activities undertaken by collaborators, finding the right time for collaboration (NTM) and managing virtual presence (DVP) can become quite difficult [14, 31]. Moreover, each person, generation or culture has a different perspective on how to use time and manage the different processes of daily life (GDR) [14, 25, 30]. It is therefore necessary to examine whether the limitations are also related to the collaborators' cultural differences. One barrier to digital collaboration can be the availability of the necessary technology [11]. With regard to iGOAL, CDP may have trouble collaborating if the investment costs for technology are high (HIC) [2, 5, 11] or if the technology is too complex (TCX) [11] due to overlaps between functionality and system design. Thus, it is necessary to distinguish barriers caused by technological experience. In the context of intergenerational and global collaboration, inequality in technological access is related to the system design, such as lack of support for different user languages and customization of the user interface as well as to access to different locations and devices (UTA) [11, 32].

Another barrier to technical and operational issues is the lack of independence or degree of freedom [1] that allows information to be presented according to users' preferences and needs (LAM). Moreover, the lack of integration of enjoyable activities into the system (ILA) can potentially increase the stress-related effects of technological use [5, 28]. All barriers explained in this section exist within the context of iGOAL. However, how CDP and non-CDP perceive these barriers has not yet been studied. Therefore, we provide an overview of the objectives of our study (see Figure 1), as well as the initial hypothesis (Ho), to uncover correlations between intergenerational barriers to cultural background:

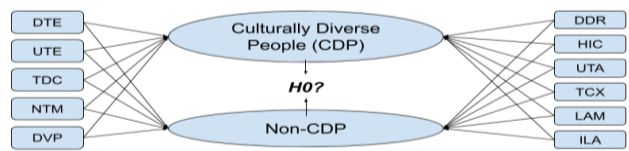

Fig. 1. Overview of the Study Objective

**Ho** = There is no difference in the perceived dimension of the technical barrier to intergenerational innovation between cultural and non-cultural migrants.

With regard to the analysis of barriers based on cultural differences, Hofstede [20] offers a country-level analysis for the cultural dimension, with a focus on the adoption of technology in multicultural collaboration [9]. The six cultural dimensions offered by Hofstede include [20]: 1) power distance or how a society deals with inequalities in collaboration. 2) The degree of individualism or reflection on whether one's self-image is determined by "I" or "We". 3) Masculinity versus femininity or preferences for taking collective action in society, whether it be more toward an assertive demeanor or for the purpose of achieving a consensus. 4) An uncertainty avoidance index reflects how society reacts to unknown future situations. 5) Long-term orientation deals with maintaining some connections with the past while coping with present and future challenges. 6) Pleasure versus restraint or a society that freely allows one to satisfy their necessary and natural human urges versus a society that suppresses such satisfaction and regulates basic human urges through strict social norms. The cultural dimension of Hofstede [20] was used to analyze the cultural dimension of barriers to iGOAL.

## 3. Method

We used the quantitative method in this study [21, 43]. The quantitative method was chosen because it can be used to investigate the correlation between known variables, as well as to quantify the correlation, which can help to analyze and justify the influence of a variable on the target variable [43]. This makes it possible to test the hypothesis and determine correlations between intergenerational barriers and participants' cultural backgrounds (CDP or non-CDP). Figure 2 depicts the processes performed in this study.

In this quantitative study, we first created an online questionnaire (in English) based on the collected intergenerational barriers presented in Section 2. The questionnaire was reviewed by an English expert before being distributed. The questionnaire in this study covers the participants' demographics, regardless of age, place of residence and cultural background. In addition to demographic questions, we also created narratives in which the respondents were presented with situations in which they would need to work across generations to answer questions based on the age classification of the group. First group: senior adults (age > 40 years old) and the second group for younger adult (17 < age < 41). Respondents were then asked a question regarding their perception of barriers to work with different generations where age differences are at least 20 years younger (for older adults) or older (for younger adults); this question was rated using a Likert scale (1: very different – 5: very similar) [4]. The online questionnaire was distributed via Amazon Mturk to reach a global audience [32, 34] and to ensure that different cultural backgrounds were reflected on a global scale.

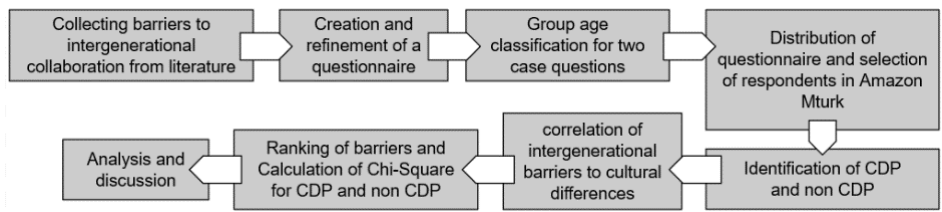

Fig. 2. Research Process

We evaluated the respondents' answers according to the experience criteria in iGOAL. We then calculated the Cronbach's alpha to determine the reliability of the data. We then established how many CDP and non-CDP participated in the study by comparing participants' cultural backgrounds with their locations of residence. The Chi-square test [26] was used to identify which barriers correlated with CDP and which did not (no correlation means that there are no differences in perception of the barriers between CDP and non-CDP). The value of the relative weight of CDP and non-CDP for each correlated barrier was then used to test H0 by calculating the value of the t-test [37] for each barrier related to the respondents' cultural conditions. H0 is rejected if the t-test value is less than the t-critical value or if the t-test value is greater than the t-critical value (two-tailed).

## 4. Result

Based on the results of the five-month questionnaire dissemination, which began in July 2018 and ended in December 2018, only 77 participants were eligible to participate. Five respondents' data were excluded from the analysis process because the respondents had no experience with iGOAL. The results of the Cronbach's alpha calculation indicate a value of 0.84, which suggests that the reliability of the data is good; therefore, the data can be used in the analysis. Table 1 shows the demographic information of the study participants.

Table 1. Demographic data of respondents

| Demographic Data | | Frequency | Percentage | Demographic Data | | Frequency | Percentage |
|---|---|---|---|---|---|---|---|
| Culturally diversity | CDP | 33 | 45,8 | Group age | Younger adult | 48 | 62 |
| | Non CDP | 39 | 54,2 | | Senior adult | 29 | 37 |
| Continent of | Africa | 2 | 2,6 | Continent of | Africa | 4 | 5,2 |

| residence | Asia | 20 | 26,0 | cultural background | Asia | 23 | 29,9 |
|---|---|---|---|---|---|---|---|
| | Europe | 2 | 2,6 | | Europe | 28 | 36,4 |
| | North America | 49 | 63,6 | | North America | 19 | 24,7 |
| | South America | 4 | 5,2 | | South America | 3 | 3,9 |
| | Australia/Oceania | 0 | 0,0 | | Australia/Oceania | 0 | 0,0 |

Furthermore, the barriers to intergenerational innovation influenced by cultural differences indicate that only 6 out of 11 technical barriers (54%) perceive the barriers differently (See Table 2). Lack of time to collaborate (NTM) is the most influential barrier to CDP.

Table 2. Descriptive Data Results from Respondents

| Technical or Operational Barriers/Challenges to IGOAL | Relative Weight (RW) | Chi-square (Chi) value | Correlation with cultural differences Chi Table = 9,49 | Comparison of RW for correlated barriers | | | |
|---|---|---|---|---|---|---|---|
| | | | | CDP | Rank | Non-CDP | Rank |
| Differences in technological experiences (DTE) | 72,2 | 6,43 | No correlation (Chi value < Chi Table) | | | | |
| Unsupportive technological environment (UTE) | 59,4 | 10,7 | Yes | 54,55 | 4 | 63,59 | 6 |
| No training available for digital collaboration (TDC) | 62,2 | 12,2 | Yes | 55,15 | 3 | 68,21 | 3 |
| No right time to collaborate (NTM) | 68,3 | 13,8 | Yes | 64,24 | 1 | 71,79 | 1 |
| Differences in daily routine (DDR) | 67,5 | 3,1 | No correlation (Chi value < Chi Table) | | | | |
| Difficulty to manage virtual presence (DVP) | 59,7 | 13,2 | Yes | 53,33 | 5 | 65,13 | 4 |
| Higher technology investment cost (HIC) | 63,3 | 5,2 | No correlation (Chi value < Chi Table) | | | | |
| Unequal in technological access (UTA) | 60,3 | 9,5 | No correlation (Chi value < Chi Table) | | | | |
| Technological complexity (TCX) | 63,3 | 9,6 | Yes | 55,76 | 2 | 69,74 | 2 |
| Lack of independence (LAM) | 55,6 | 13,1 | Yes | 45,45 | 6 | 64,10 | 5 |
| No integration of joyful activities (ILA) | 61,4 | 6,8 | No correlation (Chi value < Chi Table) | | | | |
| Calculation of t-test | | | t-test value = -4,409641211; P(T<=t) two-tail = 0,002257487; t-Critical two-tail = 2,306004135 | | | | |

Based on Table 2, the t-test for two samplings (assuming unequal variances) can be calculated by comparing the Relative Weight (RW) values of the two groups for the correlated barriers. Based on the results of the statistical calculation for the t-test, the t-test value (-4.409) is less than the t-critical two-tail (2.3) results, which means that H0 is rejected; in other words, CDP and non-CDP perceive the barriers differently. The percentage of influence of each correlated barrier to both groups, as well as the overview of the study results, can be seen in Figure 3. Next, we will discuss the results of our research and analyze the research contributions and limitations and future research proposals.

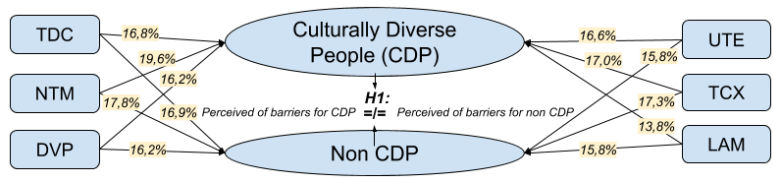

Fig. 3. Percentage of Barriers that Perceived Differently Between CDP and Non-CDP.

## 5. Discussion, Limitation and Further Research Direction

Our results suggest that six barriers correlate to intergenerational innovation, which is influenced by differences in cultural background. Moreover, other barriers are also relevant in an intergenerational context. From Table 2 it can be seen that only TDC, NTM, DVP, UTE, TCX, and LAM are perceived differently and correlated with CDP. Therefore, based on the cultural dimensions of Hofstede [20], we discuss why these correlated barriers can hinder intergenerational innovation (see Table 3). As shown in Table 3, regarding the barrier related to the availability of time to collaborate, a significant problem may arise due to differences in long-term orientation, the degree of acceptance for uncertainty and the degree of individualism, all of which are correlated with the cultural aspect.

Table 3. Discussion on Cultural Dimension

| Barriers | Correlated Cultural Dimension of Hofstede [20] |
| --- | --- |
| No right time to collaborate | 1) The dimension of individualism versus collectivism: people in a culture of individualism tend to work independently and, in this way, often avoid frequent contact with people of other generations. 2) The dimension of degree for uncertainty avoidance: people with less uncertainty avoidance tend to work spontaneously, which can lead to difficulties in arranging a meeting with people who need a fixed appointment. 3) The dimension of degree for long-term orientation: if iGOAL does not bring benefits for the future, then people with a higher score for long-term orientation would put other activities in the foreground. |
| Complexity of technology | 1) The dimension of degree for uncertainty avoidance: the higher the complexity of the technology, the lower the acceptance [38]. 2) The dimension of degree for long-term orientation: the more complex a technology is, the more uncertain is how much time and effort it takes to learn it. |
| Lack of training | 1) The dimension of degree for power distance: strategies for developing human resource skills, such as the availability of trainers, access to training, and training facilities. 2) The dimension of degree for uncertainty avoidance: uncertain about the positive impact of training on competence improvement. 3) The dimension of degree for long-term orientation: people may feel insecure about available online training, whether it will have a positive impact on improving their competence. |
| Unsupportive technological environment | 1) The dimension of degree for uncertainty avoidance: people with a low level of uncertainty - avoidance tend to quickly, adopt a system that is more familiar to them and may avoid trying new technology. 2) The dimension of degree for long-term orientation: a feeling of being on the non-supporting technological environment when people think that current technological capabilities are not relevant for working with different generations. 3) The dimension of indulgence versus restraint: technological limitations in integration with other systems can limit the possibility of combining the familiar digital tools for a particular generation in collaboration. |
| Difficulty to manage the virtual presence | 1) The dimension of individualism versus collectivism: The different attitudes and working styles towards communication and collaboration can affect how they use digital technology to collaborate. |
| Lack of independence | 1) The dimension of degree for power distance: a low power distance score tends to avoid equality in the cultural background appreciation. 2) The dimension of indulgence versus restraint: people with a higher level of indulgence may need a system that is more flexible and can be used with different types of devices to access the system. 3) The dimension of masculinity versus femininity: no appreciation for individual work in the collaboration. |

Furthermore, based on the study results, different interventions can be proposed using a developed collaborative system. We analyzed the barriers based on a project that is used to support the real-time co-authoring of Open Educational Resources (OER) [32, 33]. Although the system [33] was not specifically developed for an intergenerational setting, this study [33] was selected because the developed system includes aspects of collaborative innovation in a multicultural and multinational setting that fit our study objective. Therefore, the proposed system requirements [33] can be further developed for intergenerational collaboration by considering the six correlated barriers to intergenerational collaboration. First, in regard to the barrier "no right time to collaborate", the system designer can design a system that provides semi-synchronous collaboration, thus enabling users to share tasks and activities. Providing semi-synchronous collaboration that can be used in real-time (through users' virtual presence) or not in real-time (and thus allowing collaborators to log their information history) can help overcome this barrier to intergenerational collaboration [30]. The integration of shareable and observable activity management or task management into iGOAL can also help collaborators find out when the time is right and how much time they need to invest in working together [14, 33]. Furthermore, open communication and information about the benefits of iGOAL for innovation can also enhance both generations' motivation to arrange a real-time meeting [30, 33].

Next, people can experience technological complexity due to the focus being directed on a particular culture [10]. Moreover, bias in favor of a particular culture can also occur due to the cultural background of the system's developer [10]. To address this issue, through the OER authors project, we learned that the system designer could focus on the targeted user's current digital competence and skill [32, 33, 40]. Finding balance between the challenge and the collaborator's skills can lead to experiential flow [32]. It is also possible to reduce the complexity of the technology by training people on digital collaboration [40] in an intergenerational context. In addition, to overcome a lack of training, providing digital or online guides in the form of videos or images can help avoid misunderstandings caused by the diversity in language and culture [33]. Lack of autonomy or independence to customize the user interface is the least significant perceived barrier (RW = 55.6) according to respondents as to why they refused digital collaboration with different generations. The results indicate that the current technology allows a significant degree in flexibility in system control, such as the flexibility of language selection, customization of user interface [33] and automatic translation. Moreover, there is flexibility to access the system remotely from different devices [30, 41].

Overall, based on the study results (See Table 2), there is less of a correlation between CDP and intergenerational barriers than non-CDP (mean RW of CDP = 54.74 compare to non-CDP = 67.09). Our study supports earlier studies on global citizenship that CDP can be more open to differences [8] and thus minimize the impact of barriers to iGOAL [42]. Nevertheless, our study might not cover all important barriers that can be identified through the use of interviews; This is therefore a recommendation for the future study. Also, further research may be conducted to narrow the scope of the study on cultural diversity by focusing only on the study of differences in a multicultural country. However, we discussed the root analysis of barriers based on a higher level of the cultural dimension of Hofstede [20] that allows for a broader analysis of the barrier. In this study, we only focus on intergenerational barriers that are influenced by cultural diversity; Moreover, another study limitation is that the proposed interventions are not tested on a working prototype. Since the questionnaire is an online questionnaire, it is difficult to find senior adults that are working online in a global setting. We also collected information related to age and, therefore, as a further recommendation, the collected data can be analyzed to evaluate and identify significant barriers related to demographic information such as gender, age and place of residence. Secondly, to determine the system requirements for age differences, a positive computing approach [7, 36] can integrate a problem-based system design. The positive computing approach not only concentrates on the barriers to well-being but also supports the well-being determinants of the users and increases the human potential [7]. The approach provides principles in designing technologies that promote user wellbeing through openness of activity, difficulty and progress, automatic or playful feedback, or awards to reduce techno-stress [7, 32].

## 6. Conclusion

In summary, we have found that CDP consider six iGOAL technical barriers differently from non CDP in this study. Nevertheless, CDP tend to not consider technical barriers to be the main reason for refusing to work with different generations. Therefore, CDP can potentially be more open to digital collaboration in an intergenerational environment. More generally, our study provides insight into the critical barriers for multicultural collaborators within the context of an intergenerational environment. The outcome of this study on barriers can help to develop design requirements for tools in education, healthcare, and business, as intergenerational interactions are common in these areas.

## References

[1] Ana C. Amaro, Lídia Oliveira, and Ana I. Veloso. (2016). "Let's Build our Family Tree!". Grandparents and Grandchildren Using Tablets Together. *Procedia Computer Science* 100, 619–625.
[2] Colibaba C. Anca, Colibaba Stefan, and Petrescu Lucia, Eds. (2013). *MyStory-digital kit for story telling as a therapy*. IEEE.
[3] Jomara Binda, Hyehyun Park, John M. Carroll, Natalie Cope, Chien W. T. Yuan, and Eun K. Choe, Eds. (2017). *Intergenerational sharing of health data among family members*. ACM.

[4]  Harry N. Boone and Deborah A. Boone. (2012). "Analyzing likert data". *Journal of extension* **50 (2)**: 1–5.
[5]  Gillian M. Boulton-Lewis, Laurie Buys, Jan Lovie-Kitchin, Karen Barnett, and L. N. David. (2007). Ageing, learning, and computer technology in Australia. *Educational Gerontology* **33 (3)**: 253–270.
[6]  Karla Brücknerová and Petr Novotný. (2017). "Intergenerational learning among teachers. Overt and covert forms of continuing professional development". *Professional development in education* **43 (3)**: 397–415.
[7]  Rafael A. Calvo and Dorian Peters. (2014). *Positive computing. Technology for wellbeing and human potential*. MIT Press.
[8]  April Carter. (2013). *The political theory of global citizenship*. Routledge.
[9]  Stacey L. Connaughton and Marissa Shuffler. (2007). "Multinational and multicultural distributed teams. A review and future agenda". *Small group research* **38 (3)**: 387–412.
[10] A. Cooper, R. Reimann, D. Cronin, and C. Noessel. (2014). "About Face. The essentials of user interface design". *John Wiley & Sons*.
[11] M. K. Cresci, Hossein N. Yarandi, and Roger W. Morrell. (2010). "Pro-nets versus no-nets. Differences in urban older adults' predilections for internet use". *Educational Gerontology* **36 (6)**: 500–520.
[12] Frédéric Docquier and Abdeslam Marfouk. (2006). "International migration by education attainment", 1990–2000. *International migration, remittances and the brain drain*, 151–199.
[13] Linda F. Edelman, Tatiana Manolova, Galina Shirokova, and Tatyana Tsukanova. (2016). "The impact of family support on young entrepreneurs' start-up activities". *Journal of business venturing* **31 (4)**: 428–448.
[14] Karen Edge. (2014). "A review of the empirical generations at work research. Implications for school leaders and future research". *School Leadership & Management* **34 (2)**: 136–155.
[15] Forbes. (2011). *Global Diversity and Dinclusion Fostering Innovation Through a Diverse Workforce* (2011). Retrieved February 19, 2019 from https://i.forbesimg.com/forbesinsights/StudyPDFs/Innovation_Through_Diversity.pdf.
[16] Yuiko Fujita. (2009). *Cultural migrants from Japan. Youth, media, and migration in New York and London*. Lexington Books.
[17] Pamela A. Gordon. (2018). "Age Diversity in the Workplace". In *Diversity and Inclusion in the Global Workplace*. Springer, 31–47.
[18] Jean-François Harvey. (2012). "Managing organizational memory with intergenerational knowledge transfer". *Journal of Knowledge Management* **16 (3)**: 400–417.
[19] Donald R. Hillman. (2014). "Understanding multigenerational work-value conflict resolution". *Journal of Workplace Behavioral Health* **29 (3)**: 240–257.
[20] Geert Hofstede. (2011). "Dimensionalizing cultures. The Hofstede model in context". *Online readings in psychology and culture* 2, 1, 8.
[21] Geert Hofstede, Bram Neuijen, Denise D. Ohayv, and Geert Sanders. (1990). "Measuring organizational cultures: A qualitative and quantitative study across twenty cases". *Administrative science quarterly*, 286–316.
[22] Roma S. H. M. Icenogle. (2001). "Preparing for an age-diverse workforce. Intergenerational service-learning in social gerontology and business curricula". *Educational Gerontology* **27 (1)**: 49–70.
[23] Yong M. Kow, Jing Wen, and Yunan Chen, Eds. (2012). *Designing online games for real-life relationships. Examining QQ farm in intergenerational play*. ACM.
[24] Reginald A. Litz. (2010). Jamming across the generations. Creative intergenerational collaboration in the Marsalis family. *Journal of Family Business Strategy* **1 (4)**: 185–199.
[25] Reginald A. Litz and Robert F. Kleysen. (2001). "Your old men shall dream dreams, your young men shall see visions. Toward a theory of family firm innovation with help from the Brubeck family". *Family Business Review* **14 (4)**: 335–351.
[26] Nathan Mantel. (1963). "Chi-square tests with one degree of freedom; extensions of the Mantel-Haenszel procedure". *Journal of the American Statistical Association* **58 (303)**: 690–700.
[27] Douglas S. Massey, Joaquin Arango, Graeme Hugo, Ali Kouaouci, Adela Pellegrino, and J. E. Taylor. (1993). "Theories of international migration. A review and appraisal". *Population and Development Review* **19 (3)**: 431–466.
[28] Elizabeth Mestheneos and Alexandra Withnall. (2016). "Ageing, learning and health. Making connections". *International Journal of Lifelong Education* **35 (5)**: 522–536.
[29] Danny Miller, Lloyd Steier, and Isabelle Le Breton-Miller. (2003). "Lost in time. Intergenerational succession, change, and failure in family business". *Journal of business venturing* **18 (4):** 513–531.
[30] Diego Muñoz, Raymundo Cornejo, Francisco J. Gutierrez, Jesús Favela, Sergio F. Ochoa, and Mónica Tentori. (2015). "A social cloud-based tool to deal with time and media mismatch of intergenerational family communication". *Future Generation Computer Systems* 53, 140–151.
[31] Mihaela Nedelcu. (2017). "Transnational grandparenting in the digital age. Mediated co-presence and childcare in the case of Romanian migrants in Switzerland and Canada". *European journal of ageing* **14 (4)**: 375–383.
[32] Irawan Nurhas, Thomas de Fries, Stefan Geisler, and Jan Pawlowski, Eds. (2018). *Positive Computing as Paradigm to Overcome Barriers to Global Co-authoring of Open Educational Resources*. IEEE.
[33] Irawan Nurhas, Jan M. Pawlowski, Marc Jansen, and Julia Stoffregen, Eds. (2016). *OERauthors. Requirements for collaborative OER authoring tools in global settings*. Springer.
[34] Gabriele Paolacci, Jesse Chandler, and Panagiotis G. Ipeirotis. (2010). "Running experiments on amazon mechanical turk".
[35] Jan M. Pawlowski, Ed. (2013). *Towards Born-Global Innovation. The Role of Knowledge Management and Social Software*. Academic Conferences International Limited.
[36] Jan M. Pawlowski, Sabrina C. Eimler, Marc Jansen, Julia Stoffregen, Stefan Geisler, Oliver Koch, Gordon Müller, and Uwe Handmann. (2015). "Positive computing". *Business & Information Systems Engineering* **57 (6)**: 405–408.
[37] Graeme D. Ruxton. (2006). "The unequal variance t-test is an underused alternative to Student's t-test and the Mann–Whitney U test". *Behavioral Ecology* **17 (4)**: 688–690.
[38] Louis G. Tornatzky and Katherine J. Klein. (1982). "Innovation characteristics and innovation adoption-implementation. A meta-analysis of findings". *IEEE Transactions on engineering management*, 1, 28–45.
[39] United Nations. 2017. *World Population Prospects. Demographic Profiles* (2017). Retrieved February 19, 2019 from https://esa.un.org/unpd/wpp/Publications/Files/WPP2017_Volume-II-Demographic-Profiles.pdf.
[40] Franziska Voß, Thomas de Fries, Sabine Möbs, Jan M. Pawlowski, Celina Raffl, and Julia Stoffregen, Eds. (2018). *A Competence*

*Framework for Open Educational Resources*. The Case of the Public Sector. Springer.
[41] Greg Walsh, Allison Druin, Mona L. Guha, Elizabeth Bonsignore, Elizabeth Foss, Jason C. Yip, Evan Golub, Tamara Clegg, Quincy Brown, and Robin Brewer, Eds. (2012). *DisCo. A co-design online tool for asynchronous distributed child and adult design partners*. ACM.
[42] Matthias Wolf, Mario kleindienst, Christian Ramsauer, Clemens Zieler, and Ernst Winter. (2018). "Current and Future Industrial Challenges. Demographic Change and Measures for Elderly Workers in Industry 4.0". *Annals of the Faculty of Engineering Hunedoara-International Journal of Engineering* **16 (1)**: 67–76.
[43] Kevin B. Wright. (2005). "Researching Internet-based populations. Advantages and disadvantages of online survey research, online questionnaire authoring software packages, and web survey services". *Journal of computer-mediated communication,* **10 (3)**: JCMC1034.